\documentclass[sigconf]{acmart}
\usepackage{tabularx}
\usepackage{enumitem}
\usepackage{booktabs}  
\usepackage{graphicx}  
\usepackage{enumitem}  
\usepackage{rotating} 
\AtBeginDocument{%
  }

\setcopyright{acmlicensed}
\copyrightyear{}
\acmYear{}
\acmDOI{10.1145/3795766.3799762}
\acmConference[WebSci '26]{Proceedings of the ACM Web Science Conference 2026}
\acmISBN{}

\copyrightyear{2026}
\acmYear{2026}
\setcopyright{cc}
\setcctype{by-nc-nd}
\acmConference[WebSci '26]{18th ACM Web Science Conference}{May 26--29, 2026}{Braunschweig, Germany}
\acmBooktitle{18th ACM Web Science Conference (WebSci '26), May 26--29, 2026, Braunschweig, Germany}
\acmDOI{10.1145/3795766.3799762}
\acmISBN{979-8-4007-2504-3/2026/05}

\begin{document}

\title{Gendered Communication Patterns of Political Elites on Truth Social}


\author{Tom Bidewell}
\affiliation{%
  \institution{School of Informatics, University of Edinburgh}
  \country{UK}
}
\email{T.Bidewell@sms.ed.ac.uk}
\authornote{Both authors contributed equally to this research.}

\author{Artemis Deligianni}
\affiliation{%
  \institution{School of Informatics, University of Edinburgh}
  \country{UK}}
  \email{A.Deligianni@sms.ed.ac.uk}
\authornotemark[1]

\author{Tuğrulcan Elmas}
\affiliation{%
  \institution{School of Informatics, University of Edinburgh}
  \country{UK}
}

\author{Clare Llewellyn}
\affiliation{%
  \institution{School of Informatics, University of Edinburgh}
  \country{UK}
}

\author{Bj{\"o}rn Ross}
\affiliation{%
  \institution{School of Informatics, University of Edinburgh}
  \country{UK}
}



\begin{abstract}
  The influence of gender on online political communication remains contested, with existing scholarship providing mixed evidence as to whether gender shapes political messaging in digital environments. However, this debate has largely centred on mainstream platforms such as X (formerly Twitter), leaving the dynamics of alt-tech social media underexamined. This paper addresses this gap by analysing gendered patterns of political communication on Truth Social, a hyper-partisan platform that functions as a hub for the most committed followers of the American far right, a community closely associated with hegemonic masculine norms. To address this gap, we present the first large-scale analysis of political elite communication on Truth Social, using a novel dataset of 107k posts from 129 U.S. political figures. We examine the extent to which gender influences rhetorical style, topic framing, and audience engagement. We find that many gendered communication patterns documented on mainstream platforms persist on Truth Social. In particular, women political elites tend to express more joy and less anger than men and receive significantly higher levels of audience engagement. At the same time, more nuanced differences emerge. Although men and women political elites discuss largely similar conservative themes, they differ in how these issues are framed and in the rhetorical strategies employed. Notably, posts associated with women political elites contain higher levels of fear-based rhetoric, potentially suggesting selective adaptation in communicative style to navigate gender norms on the platform. These findings suggest that on Truth Social, an alt-tech platform with distinct ideological characteristics, mainstream gendered constraints persist, but are expressed through platform-specific communicative patterns shaped by its partisan orientation and sociotechnical environment.

  
\end{abstract}


\maketitle

\section{Introduction}
Social media is now a fundamental instrument for political communication, providing a direct channel for political elites to engage with and mobilise support \cite{mcnair2017introduction}. Within this digital landscape, the influence of gender on communication strategies has become a central focus of scholarship \cite{Russell03072023, Oden04052025, Russell02092021, storme2025, Yarchi03072018}. However, despite extensive study, the literature has not reached a consensus regarding the role of gender in online political communication. While some scholarship suggests that gender exerts minimal influence over political messaging, citing no variance in posting activity or linguistic style \cite{storme2025, Yarchi03072018}, other work identifies distinct disparities regarding thematic focus and the deployment of specific rhetorical strategies \cite{Russell02092021, Russell03072023}.

A critical limitation of this existing body of work is its overwhelming reliance on mainstream platforms such as X (formerly Twitter). These platforms operate under specific content moderation policies and host heterogeneous audiences, variables that could shape political expression. Consequently, the dynamics of gendered communication on alt-tech social media, platforms explicitly designed to bypass mainstream norms and consolidate partisan communities, remain largely underexplored. This represents a significant gap in our understanding of political communication on sociotechnical systems: we do not yet know whether the gendered dynamics observed on mainstream sites hold true in unmoderated, hyper-partisan echo chambers.

This work addresses this gap by investigating the communication patterns of political elites on Truth Social. Established by Donald Trump in 2022 \cite{Gerard_Botzer_Weninger_2023}, Truth Social offers a unique, high-stakes environment for examining the influence of gender. As an alt-tech platform, Truth Social positions itself as a haven for “free speech”, potentially altering the linguistic incentives for political actors compared with moderated sites. In addition, the user base leans heavily conservative, with high levels of support for Trump \cite{Gerard_Botzer_Weninger_2023}. Research indicates that this demographic correlates strongly with traditionalist views on gender roles \cite{vescio2021}, potentially creating pressure for posts by women political elites to diverge from those of their male counterparts. Finally, the site functions as a centralised “mouthpiece” for Donald Trump \cite{Gerard_Botzer_Weninger_2023}. This centralisation is critical given that alt-tech platforms have historically shifted from passive discourse to active mobilisation, exemplified by the January 6th Capitol insurrection \cite{jan_6_alt_tech}. Consequently, the rhetoric and gendered performance on Truth Social are not merely digital artefacts, but potential drivers of real-world political action.

To understand how gender operates within this distinct sociotechnical environment, we ask the following research questions:
\begin{itemize}[topsep=0pt, partopsep=0pt, parsep=0pt]
    \item \textbf{RQ1:} How does the gender of the political elite impact the content of the post?
    \item \textbf{RQ2:} How does the gender of the political elite impact the rhetorical style of the post?
    \item \textbf{RQ3:} How do the gender of the political elite and features of the post impact audience responses (upvotes, replies)?
\end{itemize}
We find that gendered patterns of elite political communication on Truth Social largely mirror those observed on mainstream social media platforms, despite the platform’s ideological homogeneity and association with hyper-masculine political norms. At the same time, we do identify subtle differences in how political content is framed and rhetorically articulated that may stem from the platform itself. To the best of our knowledge, this work represents the first large-scale analysis of political elite communication on Truth Social. By examining this understudied but influential platform, we expand the literature on gendered political communication.

\section{Related Work}
Scholarship on political communication in mainstream social media has produced mixed evidence regarding the role of gender, with no clear consensus on whether men and women political elites adopt distinct communicative strategies. One strand of research identifies persistent stylistic differences; for example, Russell et al. find that Republican women employ significantly more expressions of joy and fewer angry appeals than Republican men on social media, potentially to navigate gendered expectations around aggression \cite{Russell03072023}. This aligns with work by Bauer et al., who show that women candidates must strategically balance masculine and feminine stereotypes \cite{bauer_et_al}, and with McGregor et al., who demonstrate that women candidates selectively present themselves in maternal or caregiving roles on social media when doing so is advantageous~\cite{McGregor01022017}.

In contrast, other studies suggest that the digital gender gap may be narrowing. Russell, for instance, finds that men Senators are just as likely as women Senators to address women’s issues on social media \cite{Russell02092021}, while Oden et al. report no significant differences in how men and women legislators foreground women's issues in their agenda-building, even after controlling for political characteristics~\cite{Oden04052025}.

Despite the breadth of this literature, it remains overwhelmingly focused on mainstream, moderated platforms, leaving alt-tech environments comparatively underexamined. Although scholarship on Truth Social is beginning to emerge, existing research has not centrally examined the communicative practices of political elites on the platform. Gerard et al. introduced a large-scale dataset of Truth Social users and provided initial descriptive analyses of platform activity \cite{Gerard_Botzer_Weninger_2023}, while Shah et al. focused specifically on discourse surrounding the 2024 U.S. presidential election \cite{shah2024unfilteredconversationsdataset2024}. Other strands of work have examined Donald Trump’s use of Truth Social and its role in driving media attention \cite{zhang2025trump}, as well as the platform’s broader contribution to the circulation of politically radical and polarising content \cite{goldstein2024truthsocial}. Complementing this, research on cross-platform narrative migration shows that narratives originating on Truth Social employ more fear-oriented rhetoric than comparable content on X \cite{gerard2025bridgingnarrativedividecrossplatform}, highlighting the platform’s importance as a generator of distinctive and potentially influential political narratives that warrant closer examination. Wang et al. found that Truth Social is more grievance oriented but less toxic than conservative communities on Reddit~\cite{wang2026grievance}. Together, these observations highlight the need for closer analysis of elite communication on the platform. Our study therefore investigates both the prevalence of fear-based rhetoric in posts by political elites and the extent to which such rhetoric influences audience engagement within this distinctive ecosystem. 

Investigating gender on Truth Social also presents a theoretically different challenge from analysing gender dynamics on mainstream platforms due to the site's highly homogeneous user base, composed largely of strong supporters of Donald Trump and the American far right. Vescio and Schermerhorn show that support for Trump is strongly predicted by adherence to hegemonic masculinity, a worldview that promotes male dominance while devaluing femininity \cite{vescio2021}. Complementary work by Ratliff et al. demonstrates that positive attitudes towards Trump are associated with higher levels of hostile sexism \cite{ratliff2019engendering}, defined as negative attitudes and stereotypes about women \cite{glick2001ambivalent}. As a result, Truth Social functions as a distinctive sociotechnical environment in which women political elites must navigate an audience that tends to associate political legitimacy with hyper-masculine performance. This makes the platform an important, and previously underexamined, setting for understanding how gender shapes political communication in ideologically extreme online spaces.

\section{Data}
We constructed a comprehensive list of U.S. political elites maintaining official accounts on Truth Social as of 2025. We defined \textit{\textbf{political elite}} broadly to include not only current elected officeholders but also influential unelected figures within the Trump administration and the wider MAGA movement (e.g., Cabinet members, senior advisors). To assemble this dataset, we first compiled a comprehensive list of U.S. political actors using official government sources. This list included all members of the House of Representatives and the Senate, members of the National Governors Association, Cabinet officials, and other nationally prominent appointees occupying high-visibility roles (e.g., the White House Press Secretary). We then manually searched Truth Social to identify whether each individual maintained an account on the platform. Accounts were classified as official based on clear indicators of authenticity, including platform verification, follower counts consistent with public prominence, and professional presentation (e.g., official photographs or links to government or campaign websites). In addition, we manually coded the gender associated with each account using publicly available information from profiles or official webpages.

The final dataset comprises 129 accounts (including 2 Democrat political elites) and over 107k posts, representing the entirety of these users' activity on the platform. For every account, we collected metadata along with all available posts using the Truth Social API \cite{truthbrush2022}. Data collection took place between August 12 and August 14, 2025.

We acknowledge that elite political accounts are frequently managed by communication teams rather than the political elites themselves \cite{staffers_write_posts}. However, for the purpose of this study, we treat the account as a distinct digital persona. Since these communication teams are explicitly tasked with performing the political elite’s identity to an audience that perceives the content as coming from that individual, the gendered presentation of the account, rather than the gender of the specific staffer drafting the text, remains the valid unit of analysis for exploring rhetorical strategy and audience reception. This justification is consistent with other work on gendered political communication on social media \cite{BerasategiGenderAF}.

We also acknowledge the difficulty in operationally defining the start date of a subject's public prominence, creating a potential for the inclusion of posts that predate their status as political elites. To assess this risk, we manually reviewed the political careers of all individuals in the dataset by consulting publicly available biographical information. This verification showed that each individual had either (i) held political office or been active in politics prior to the launch of Truth Social in February 2022, or (ii) been elected to political office later that same year. Consequently, the risk of posts preceding the period in which these individuals were identifiable as political elites is minimal.

Additionally, while a sampling strategy based on 2025 incumbency omits actors who have recently left office, the scale of the dataset is sufficient to extract the generalisable trends of political communication required for our research questions.

The dataset is stored in an online repository and can be found at: \url{https://osf.io/hykur/overview}.

\section{Exploratory Data Analysis}
Our final dataset contains 107,393 posts authored by 129 U.S. political elites on Truth Social. Of these accounts, 77\% (N = 99) are men and 23\% (N = 30) are women. Women politicians are underrepresented across all political positions. 

Posting activity varies widely across users: the median number of posts per political elite is 2099, though certain high-profile figures (e.g. Donald Trump) contribute disproportionately. Notably, men political elites are more active on average, with a median of 4851 posts (Standard Deviation (SD) = 3613) compared to 1866 posts for women political elites (SD = 1432). Despite their smaller representation on the platform, women political elites maintain larger audiences. Median follower counts are 66,772 for women political elites (SD = 619671) versus 4,075 for men political elites (SD =1115905), and women political elites also follow more accounts on average. Engagement patterns further highlight gendered differences. Posts from women political elites receive more interaction across all measured metrics: a median of 19 replies (SD = 193) compared to a median of 4 replies for men (SD = 790), 85 reblogs (SD = 456) compared to 18 for men (SD = 2099), and 346 favourites for women (SD = 1655) compared to 58 for men (SD = 8540).

\section{Methods}
\subsection{Content}
To distinguish between political messaging and personal communication, we first classified the content of each post as either political or non-political (see Section B in Appendix in Supplementary Material for coding definitions). We used Llama 3.1 8B Instruct \cite{grattafiori2024llama3herdmodels} in a zero-shot setting to classify each post as political or non-political. To evaluate classification performance, a random sample of 100 posts was independently annotated by two human coders. Inter-annotator agreement, measured using Cohen’s Kappa, was 0.747. Posts with annotation disagreements were excluded, leaving 93 posts with human-labelled ground truth. The LLM achieved a macro F1 score of 0.90, indicating satisfactory performance for large-scale classification of the dataset. 

We then conducted topic modelling to examine the content of posts across gender and political categories. Specifically, we employed BERTopic \cite{grootendorst2022bertopic} to identify topics in the posts. To improve topic quality, we conducted a systematic hyperparameter search using Bayesian optimisation (via Optuna \cite{optuna_2019}) over the UMAP \cite{mcinnes2020umapuniformmanifoldapproximation}, HDBSCAN \cite{mcinnes2017hdbscan}, and topic reduction parameters. Each configuration was evaluated using the c\_v coherence score \cite{coherence_metric}, with the optimisation procedure seeking to maximise average coherence across topics. This procedure was repeated separately for each subset of the data (political vs. non-political; man vs. woman), and the best-performing configurations were used to perform the topic modelling (see Section B in the Appendix in the Supplementary Material file for breakdown of hyperparameters).

To complement this analysis and gain a deeper understanding of gendered differences, we also fine-tuned a RoBERTa classifier \cite{liu2019robertarobustlyoptimizedbert} to predict whether a post came from a man or woman political elite's account. Training employed standard hyperparameters, with further details provided in Section C in the Appendix in the Supplementary Material. The classifier achieved strong performance, with an F1 score of 0.90. We also repeated the training after excluding posts by Donald Trump given his prominence on the platform and the model still performed well, attaining an F1 score of 0.88.
To investigate the features driving the classifier’s predictions, we conducted a gradient-based saliency analysis \cite{wallace2019allennlp, han-etal-2020-explaining}. This method identifies the most influential words contributing to the predicted label, thereby surfacing differences between the two groups of posts. Specifically, for each text in the test set, we computed the gradient of the predicted label with respect to each token's embedding. We then calculated importance scores by taking the sum of the element-wise product between each token's embedding and its gradient across all embedding dimensions. Finally, we normalised the scores for each input post by dividing them by the L1 norm of the scores of all tokens within that post. For each word, we averaged the scores of its constituent tokens. The five words with the highest average absolute importance scores per post were recorded, after removing miscellaneous tokens and words such as sequence boundary tokens.

We then applied the same topic modelling procedure as above to extract salient topics from the influential words in men and women political elites’ posts. Gradient-based saliency provides direct evidence of which tokens contribute to the model’s predictions, offering greater interpretability than attention weights, which may not accurately capture token influence \cite{bastings2020elephant}.

\subsection{Rhetorical Style}
We examined the rhetorical style of political elites’ posts by extracting features for emotions present, hostile rhetoric, and various linguistic dimensions.

First, to capture the emotional profile of the posts, we used the English subset of the BRIGHTER dataset, a multilingual emotion detection from text dataset \cite{muhammad2025brighterbridginggaphumanannotated}. We fine-tuned RoBERTa-based classifiers \cite{liu2019robertarobustlyoptimizedbert} to detect five distinct emotions: anger, fear, joy, sadness, and surprise. At inference, each model was applied independently to the text, yielding multi-label predictions for each post. If no emotion was present, the text was labelled as neutral. The average F1 score across the five emotions was 0.70 (see Section B in the Appendix in the Supplementary Material for training details).

Second, to explicitly measure the hostile rhetoric prevalent in the posts, we trained classifiers for fear speech and hate speech (see Section B in the Appendix in the Supplementary Material for definitions). Using human-annotated data from Gab, a right-wing social media platform similar to Truth Social \cite{fearspeech_model}, we fine-tuned RoBERTa-based classifiers that had already been trained for hate speech detection \cite{barbieri-etal-2020-tweeteval}. This approach followed similar work on fear and hate speech detection \cite{fearspeech_model, gerard2025bridgingnarrativedividecrossplatform}. We trained two binary classifiers for the existence of fear speech and hate speech respectively. See Section B in the Appendix in the Supplementary Material file for further training details and performance metrics of these models.

Lastly, we analysed a range of linguistic features for each post, including average sentence length, the root type–token ratio (RTTR; a length-adjusted measure of lexical diversity), sentiment scores \cite{hutto2014vader}, SMOG index reading ease scores \cite{mclaughlin1969smog}, and emoji usage. 

\subsection{Audience Responses}
To answer RQ3 we wanted to test (1) which features of a post predicted higher counts of replies and upvotes, and (2) whether gender was predictive of replies and upvotes over and above the features of the posts.

\subsection{Gender and Statistical Analyses}
To account for the fact women are underrepresented, all the statistical analyses we report here are conducted on a random subset with 4000 posts from women and 4000 posts from men, to ensure both genders are represented equally. We report 95\% Bayesian credible intervals, which indicate the range we are 95\% certain the true value of our estimate falls in. 

\section{Results}
\subsection{RQ1: Content}

\textbf{Political content}
Our classifier identified the vast majority of posts as political in nature, with men posting political content 83\% of the time and women 81\% of the time. 

We fit a Bayesian Logistic Regression with the content of posts (political/non-political) being the outcome variable predicted by gender, political position and  each politician's z-scored average posts per week. The reference levels of the categorical predictors was men (gender) and House of Representatives (political position). We also tested for a possible interaction of gender and position and compared the interaction model to the additive model using Leave-One-Out validation. Although the interaction model had slightly higher LOO score ($\Delta$ELPD = 4.1, Standard Error (SE) = 2.3), the advantage is modest and within sampling uncertainty. Given the absence of strong theoretical motivation for the interaction, we present results from the additive model. Below we summarise the main findings, but the full results can be found in Section D in the Appendix in the Supplementary Material. 

Gender was predictive of whether a post was political or not political. Holding all other variables constant, women had 47\% decrease in odds of making a political post compared to men (OR=0.53, 95\% CrI[0.45, 0.61]). For every 1 Standard Deviation point increase in average posts per week the odds of a post being political increase by 40\% (OR = 1.40, 95\% CrI[1.30, 1.51]). There was a main effect of political position. Compared to politicians in the House of Representatives, posts made by the President (OR= 0.39, 95\% CrI[0.31, 0.48]), the head of the Federal Bureau of Investigation (OR= 0.40, 95\% CrI[0.31, 0.51]), Governors (OR= 0.23, 95\% CrI[0.14,0.36]) and members of the Cabinet (OR= 0.70, 95\% CrI[0.57, 0.82]) all had decreased odds of being political in content.

\textbf{Topic content} Topic modelling of political posts revealed substantial thematic overlap across genders, with both men and women political elites frequently addressing international conflicts (Israel–Palestine, Russia–Ukraine), inflation under the Biden administration, and Kamala Harris.

Gender-specific emphases also emerged. Men political elites more often discussed immigration and border security, oil and gas, and education and schooling. Women political elites, in contrast, placed greater emphasis on the Chinese Communist Party, the judiciary and court system, antisemitism on university campuses, and debates surrounding gun rights and the Second Amendment. Geographic references were especially prominent in women political elites’ posts, most notably through Kari Lake’s repeated focus on Arizona, where she narrowly lost a Senate race. 

Topic modelling of non-political posts revealed a narrower thematic range. Both men and women political elites discussed veterans and natural disasters such as hurricanes. Men political elites additionally invoked Christianity, while women political elites more frequently highlighted geographical locations, particularly Arizona, South Dakota, and Georgia. Men political elites also referenced place with discussions centred on Iowa and Pennsylvania.

\textbf{Topic framing} The gradient-based saliency analysis provided further insights into gendered patterns of political communication. Using topic modelling on the most influential words for the classifier, we observed clear differences between men and women political elites in how they discussed topics. While both groups discussed border-related concerns, they framed the topic in distinct ways. Women political elites primarily linked the issue to the fentanyl crisis and connected Biden to border policy, whereas men political elites emphasised the broader presence of ``illegal aliens”. Similarly, both groups engaged with the Israel–Palestine conflict, though women political elites focused on antisemitism while men political elites concentrated on Hamas. In addition, men political elites often referred to Democrats directly, characterising them with terms such as ``radical”, a strategy not seen among women political elites.

\subsection{RQ2: Rhetorical Style}

We fit five Bayesian Binary regressions where the outcome was the presence or absence of each emotion (fear, anger, sadness, surprise, joy) from a given post. Our predictors included gender,  political position, post content (political/non-political) and type of speech (normal/hate speech/fear speech). The reference levels are man (gender), House of Representatives (political position), Non-political (political content) and normal speech (type of speech). We tested for a possible interaction of gender with political position, but Leave-One-Out validation results indicated no important improvements in prediction for each of the 5 emotion labels ( Anger: ($\Delta$)ELPD = -1.0, SE= 0.2 ; Fear: ($\Delta$)ELPD = -2.5, SE= 1.7 ; Joy:  ($\Delta$)ELPD = -1.3, SE=  0.4 ; Sadness: ($\Delta$)ELPD = -1.8, SE=2.1 ; Surprise: ($\Delta$)ELPD = -0.3, SE= 1.6). Given we have no strong theoretical reasons to keep an interaction between gender and political position we report here the additive model, focusing only on variables which were shown to be predictive of the presence of a given emotion in a post. The full results table can be found in Section D in the Appendix in the Supplementary Material. 

\begin{figure}[htbp] 
    \centering
    {\raggedright
    \hspace*{-0.5cm}%
    \includegraphics[width=0.5\textwidth]{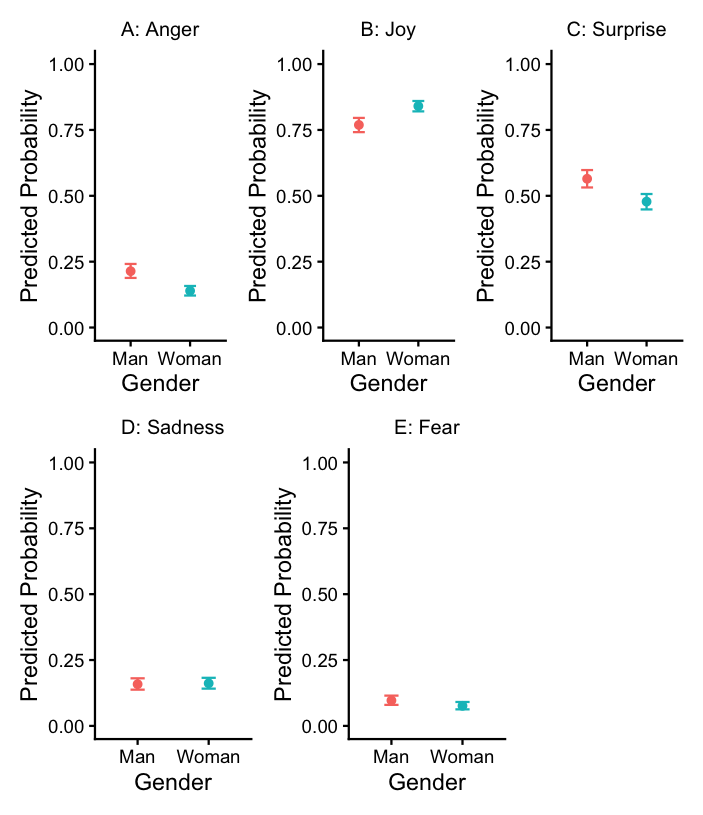}
    \par}
    \caption{Predicted probabilities of using each of the 5 emotionsfor men and women political elites. The error bars indicate 95\% Credible Intervals. Each sub-figure represents the estimated gender probabilities of using that emotion based on the respective Logistic Regression model fitted. For sadness and fear, we do not find evidence gender predicting their use in posts (Sub-figures D and E).}
    \label{fig:emotions_combined}
\end{figure}

\textbf{Anger in Posts} Gender was predictive of the use of anger in a given post. Holding all other variables constant, the odds of women using anger in their posts was 41\% less compared to men (OR= 0.59, (95\% CrI [ 0.52, 0.68] (See Subfigure A Figure \ref{fig:emotions_combined}). The odds of a post containing anger were increased by 1970\% for political posts compared to non-political posts (OR=20.70, 95\% CrI [17.81, 24.05]). Similar results were found for fear speech and hate speech. The odds of a post containing anger increased by 767\% if it also contained fear speech (OR=8.67, 95\% CrI [5.42, 14.59]) or by 366\% if it contained hate speech (OR= 4.66, 95\% CrI [3.06, 7.32]), when compared to normal speech. Compared to the House of Representatives, politicians in the Senate (OR= 0.65, 95\% CrI [0.52, 0.82]), Governors (OR= 0.47 , 95\% CrI [0.26, 0.84]), and the President had decreased odds of using anger in their posts (OR= 0.73, 95\% CrI [0.60, 0.90]). 



\textbf{Fear in Posts} Gender was predictive of the use of fear in a given post. Holding all other variables constant, there was a 23\% decrease in the odds of women using fear in their posts compared to men (OR= 0.77, 95\% CrI [0.68, 0.88]) (Probabilities visualised in sub-figure E in Figure \ref{fig:emotions_combined}). Political posts had a 156\% increase in odds of containing fear compared to non-political posts made by men (OR= 2.56, 95\% CrI [2.14, 3.13]). The odds of a post containing fear when it also contained fear speech increased by 589\% (OR= 6.89, 95\% CrI [5.59, 8.41]) and by 186\% if it contained hate speech (OR=2.86, 95\% CrI[2.25, 3.56]) compared to normal speech. Compared to politicians in the House of Representatives, the head of the Federal Bureau of Investigation (OR= 0.54, 95\% CrI[0.41, 0,71]) and the President had decreased odds of using fear in their posts (OR= 0.81, 95\% CrI[0.66, 0.99]). 

\textbf{Joy in Posts} Gender was predictive of the use of joy in a given post. Holding all other variables constant, the odds of using joy in a post increased by 58\% for women compared to men (OR= 1.58, 95\% CrI[1.42, 1.77]) (See sub-figure B in Figure \ref{fig:emotions_combined} for the visualised proabablities). Compared to non-political posts, political posts had a decrease of 82\% in the odds of containing joy (OR= 0.18, 95\% CrI[0.15, 0.20]). Similar patterns were found for posts containing either fear speech or hate speech. The odds of containing joy decreased by 80\% for posts containing fearpseech (OR = 0.20, 95\% CrI[0.15, 0.26]) and by 61\% for posts containing hate speech (OR= 0.39, 95\% CrI[0.30, 0.50]). There was evidence that for some political positions the odds of using joy in non-political posts were higher. Compared to the politicians in the House of Representatives, those in the Cabinet (OR= 1.36 , 95\% CrI [1.13, 1.63]), the Senate (OR= 1.42, 95\% CrI [1.61, 1.73]), Governors (OR= 1.95 , 95\% CrI [1.15, 3.42]), the head of Federal Bureau of Investigation (OR= 1.45, 95\% CrI [1.17, 1.77]), the Press Secretary (OR= 2.03, 95\% CrI [1.39, 3.00]) and the President (OR= 1.51, 95\% CrI [1.30, 1.79]) had higher odds of using joy in their posts.

\textbf{Sadness in Posts} When controlling for all other variables, gender was not predictive of whether a post contained sadness or not (OR= 1.03, 95\% CrI[0.92, 1.15]) (See sub-figure D in Figure \ref{fig:emotions_combined} for visualised probabilities). There was a 278\% increase in odds of a post containing sadness if it was political, compared to non-political (OR=3.78, 95\% CrI[3.25, 4.39]). Compared to normal speech, the odds of a post containing sadness if it also contained fear speech increased by 1002\% (OR=  11.02,95\% CrI[8.08, 15.03]) or by 210\% if it contained hate speech (OR=3.10 ,95\% CrI[2.48,3.94]). We also observe that compared to the House of Representatives, those in the Senate (OR= 0.75, 95\% CrI[0.60, 0.92]), the head of the Federal Bureau of Investigation (OR= 0.45, 95\% CrI[0.35, 0.57]), the Press Secretary (OR= 0.43, 95\% CrI[0.28, 0.65]), and the President (OR= 0.81, 95\% CrI[0.69, 0.95]) have decreased odds of making a post containing sadness.

\textbf{Surprise in Posts} When controlling for all other variables, women had a 30\% decrease in the odds of using surprise in their posts compared to men (OR= 0.70 ,95\% CrI[0.63, 0.79]) (See sub-fgure C in Figure \ref{fig:emotions_combined}, for the visualised probabilities). Comparing political posts to non-political posts political posts had an 156\% increase in odds of containing surprise (OR=2.56 ,95\% CrI[2.27, 2.86]). Posts which contained fear speech (OR=2.18,95\% CrI[1.68, 2.83]) or hate speech (OR= 2.83 ,95\% CrI[2.05, 3.97 ]) had increased odds of containing surprise compared to normal speech; 156\% and 183\% increase in odds respectively. Compared to politicians in the House of representatives, members of the Senate (OR= 0.59 , 95\% CrI [0.48, 0.73]) and the Vice President (OR= 0.51, 95\% CrI [0.26, 0.97]) had decreased odds of using surprise in their posts, whereas the Head of the Federal Bureau of Investigation (OR= 2.44 ,95\% CrI[1.88, 3.25]) and the President (OR= 1.93,95\% CrI[1.60, 2.34]) had a 144\% and 93\% increase respectively in the odds of using surprise in their posts. 

\textbf{Fear speech and Hate speech}
We fit a Bayesian Multinomial regression where type of speech (normal, fear speech, hate speech) is predicted by gender, political position and politician's z-scored weekly average posts. The reference levels of the categorical predictors are men (gender), in the House of Representatives (political position). Similar to the political content analyses, we also tested for a possible interaction of gender and position and compared the interaction model to the additive model. Although the interaction model had slightly higher LOO score (($\Delta$)ELPD = 1.0, SE =  1.5), the advantage is modest and within sampling uncertainty. Given the absence of strong theoretical motivation for the interaction, we present results from the additive model. Below we summarise the main findings, and the full results can be found in Section D of the Appendix in the Supplementary Material.

Gender was predictive of the use of fear speech but not hate speech, relative to normal speech. Compared to men, women have a 49\% increase in odds of using fear speech in their posts relative to normal speech (OR= 1.49, 95\% CrI[1.17, 1.90]). Conversely, gender does not predict a difference in the odds of using hate speech in posts between men and women, compared to normal speech (OR= 0.95, 95\% CrI[0.72, 1.26]). The average posts per week Edinburghpredicted politicians' odds of using hate speech (OR= 0.80, 95\% CrI[0.70, 0.93]) but not fear speech (OR= 1.00, 95\% CrI[0.90, 1.13]) relative to normal speech. In other words for every 1 Standard Deviation point increase in the average posts per week there was a 20\% decrease in the odds of using hate speech, but there is no evidence of a similar association for fear speech.

Additionally, political position predicted the use of fear speech. Compared to the House of Representatives, members of the Senate (OR= 0.58 , 95\% CrI[0.34, 0.94]), the head of the Federal Bureau of Investigation (OR= 0.34, 95\% CrI[0.16 , 0.66]), the Press Secretary (OR= 0.26 , 95\% CrI[0.07, 0.78] ) and the President (OR= 0.36 , 95\% CrI[1.23, 0.62]) had decreased odds of using fear speech in their posts.

Political position also predicted the use of hate speech. Compared to politicians in the House of Representatives, those in the Cabinet posts (OR =  0.58, 95\% CrI[0.35, 0.92]) and the Press Secretary posts (OR = 0.32, 95\% CrI[0.09, 0.98]) had decreased odds of using hate speech in their posts, whereas the Head of the Federal Bureau of Investigation had slightly increased odds of using hate speech on his posts (OR = 1.62 , 95\% CrI[1.07, 2.39]). However, there credible intervals around the Press Secretary’s estimate span a very large and very small values, with the upper bound being very close an Odds Ratio of 1, so this should be interpreted with caution as the true value could be quite small.

 \textbf{Linguistic features}: We fit 3 models, to investigate whether gender predicts linguistic measures like VADER Compound Score, RTTR (Root of Type to Token Ratio) and number of sentences politicians use in their posts. For the VADER Compound Score and RTTR we used a gaussian regression, whilst for the count of sentences we used a Poisson regression. Our predictor variables for all 3 models included gender, position, political content, hate speech and z-scored average posts per week, as well as random intercept for each individual politician to account for variation in individual's linguistic habits. 

 \textbf{Sentiment Analysis} To examine the sentiment of posts, we used the VADER Compound Score \cite{hutto2014vader}. The compound score provides a normalised measure for overall sentiment of a text, ranging from -1 (most negative) to +1 (most positive), with values around 0 indicating neutral sentiment. In terms of sentiment score, we also do not find a difference between men and women (\(\beta\) = 0.06, 95\% CrI[-0.03, 0.15]). For ever 1 Standard Deviation point increase in the politicians’ average posts per week, we observe a very small increase in the VADER Compound Score of the post (meaning more positive affect) (\(\beta\) =0.02, 95\% CrI[0.01, 0.04]). However, as the lower bound is very close to 0, this result should be interpreted with caution. Compared to politicians in the House of Representatives, we find governors have a 0.27 points higher VADER Compound score in their posts (\(\beta\)= 0.27, 95\% CrI[0.03, 0.49]). In terms of post content, we observe that political posts have a decrease of 0.34 points in VADER Compound Score compared to non-political posts (\(\beta\) = -0.34, 95\% CrI[-0.37, -0.31]). Controlling for political content, we find that when posts contain fear speech there is on average a 0.57 point decrease in VADER Compound Score (\(\beta\)= -0.57, 95\% CrI[-0.62, -0.51]), and for posts containing hate speech a decrease of 0.47 points (\(\beta\)= -0.47, 95\% CrI[-0.53, -0.41]), compared to normal speech posts. 

\textbf{Root Type-Token Ratio} For RTTR, we find that women have a 0.34 point increase in vocabulary diversity in their posts compared to men (\(\beta\)=  0.34, 95\% CrI[0.02, 0.66]). However, the lower bound credible interval is very close to 0, so this result should be interpreted with caution. We also observe a very small association for politician’s average posts per week; for every 1 SD increase in average posts per week there in a small decrease of -0.06 points in vocabulary diversity (\(\beta\)= -0.06, 95\% CrI[-0.10 , -0.01]). Because the upper bound is very proximate to 0, this result should be interpreted with caution. Political posts had an increase of 1.23 points in vocabulary complexity compared to non-political posts (\(\beta\)= 1.23, 95\% CrI[ 1.15, 1.31]). Lastly, we find that posts which contained fear speech were predicted to have 0.59 points increase in vocabulary complexity compared to normal posts (\(\beta\)= 0.59, 95\% CrI[0.47, 0.72]), but we do not find evidence of an association for hate speech posts (\(\beta\)=0.13, 95\% CrI[-0.01, 0.27]).

\textbf{Number of sentences} We do not find that women differed from men politicians in the expected counts of sentences in their posts, when holding all other variables constant (RR = 1.03, 95\% CrI[ 0.89, 1.20]).  Furthermore, we find a main association of political position. Compared to politicians in the House of Representatives, we find that the Border Tsar, Tom Homan, a senior immigration enforcement official appointed by Donald Trump, had an increase of 170\% in the expected count of sentences in posts (RR = 2.70, 95\% CrI[1.54, 4.58]). Additionally, we find that political posts are predicted to have 25\% increase in the expected counts of sentences compared to non-political posts (RR = 1.25, 95\% CrI[1.85,1.30]). Posts containing fear speech had an 17\% increase in expected counts of sentences compared to posts using normal speech (OR = 1.17, 95\% CrI[1.09, 1.25]), though a similar association was not observed for hate speech posts (RR = 1.03, 95\% CrI[0.95, 1.11]). 

\subsection{RQ3: Audience Responses}

We fit two separate Bayesian negative binomial regression models predicting (1) reply count and (2) upvotes count. We analyse post replies and upvotes to glean the audience response as a function of the politician’s gender, position, and z-scored average posts per week, as well as the post’s content and type of speech. The reference levels for the categorical predictors are as follows: men (gender), House of Representatives (position), non-political (content) and normal speech (type of speech). Because there was moderate co-linearity between all the 5 emotion measures, we include a z-scored measure of emotion, that is a count of how many emotions out of 5 (anger, fear, sadness, joy, surprise) are present in a post, centered around the mean. Additionally, we tested for an interaction between gender and position. When comparing the interaction model with the additive model we find that the interaction improves model prediction meaningfully for replies (($\Delta$)ELPD=  135.2, SD= 16.0) and for upvotes (($\Delta$)ELPD=  153.1 , SD= 18.8). Hence, we report the interaction model results below, focusing on variables which predicted replies and upvotes. The full results can be found in Section D of the Appendix in the Supplementary Material.

\begin{figure}[htbp] 
    \centering
    {\raggedright
    \hspace*{-0.6cm}%
    \includegraphics[width=0.40\textwidth]{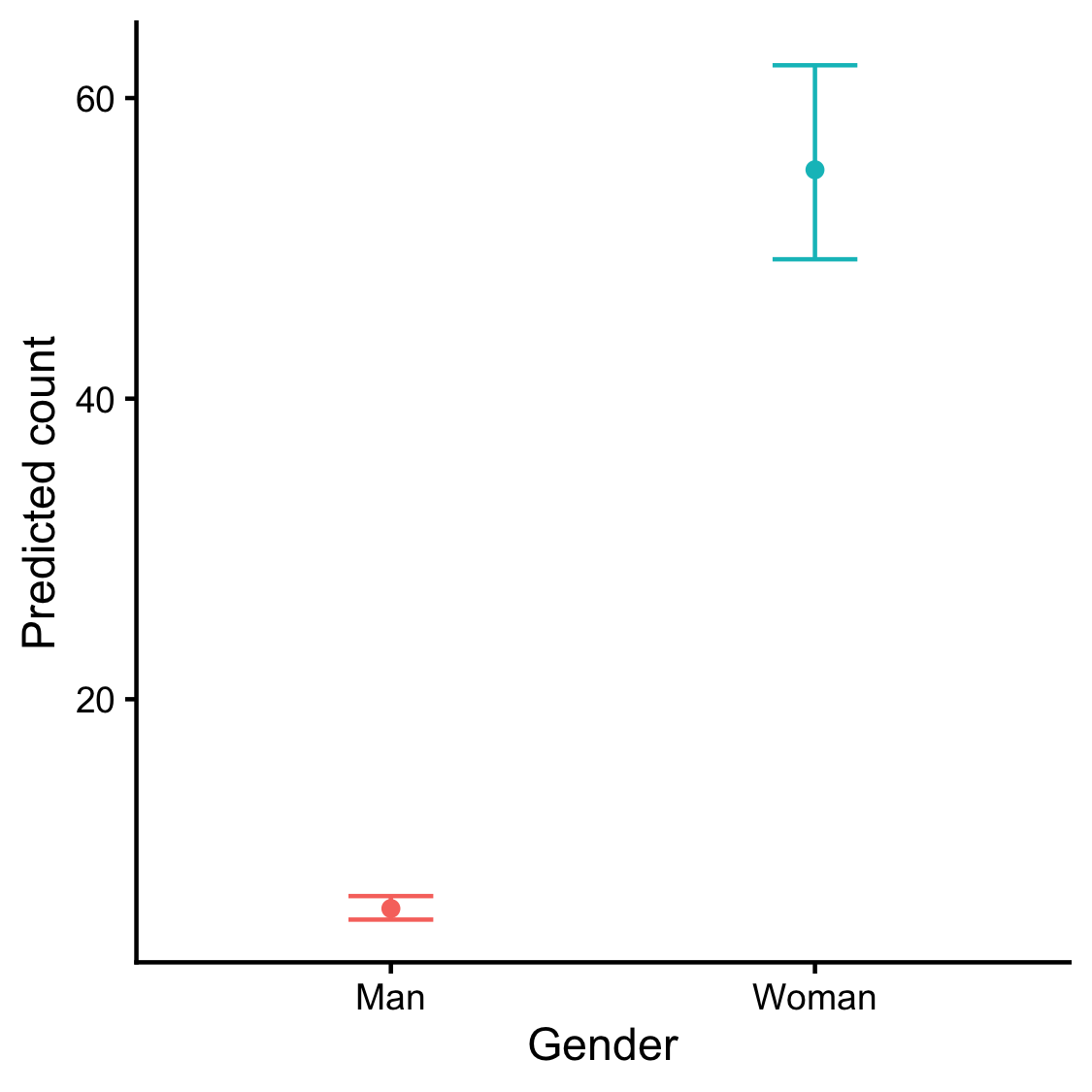}
    \par}
    \caption{Estimated counts of replies received by men and by women political elites to their posts when controlling for covariates. Errorbars indicate 95\% Credible Intervals.}
    \label{fig:replies_gender_main}
\end{figure}

\begin{figure}[htbp] 
    \centering
    {\raggedright
    \hspace*{-0.5cm}%
    \includegraphics[width=0.5\textwidth]{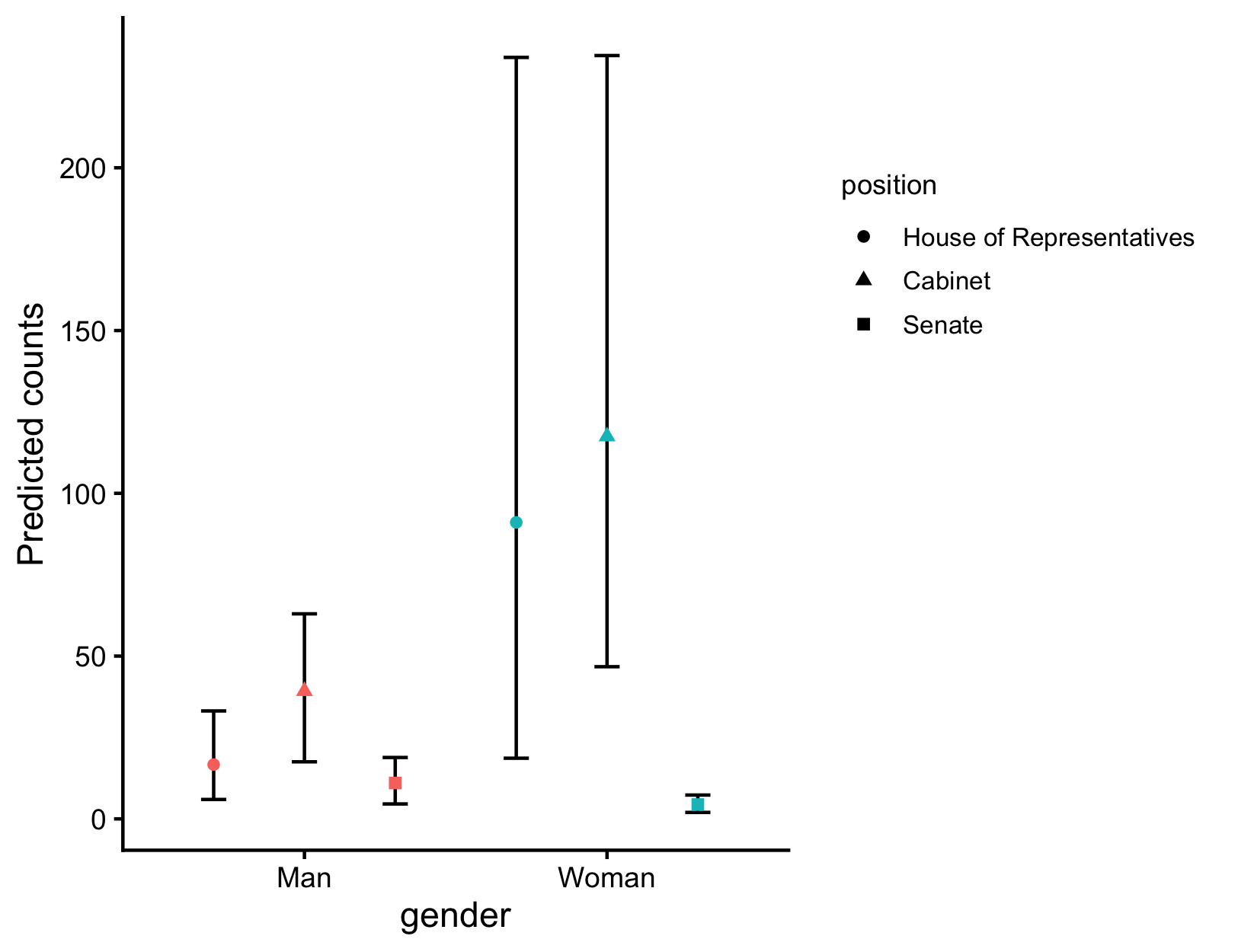}
    \par}
    \caption{Interaction of gender and political position for reply counts. Estimates are predicted counts of replies received by posts of political elites depending on their gender and political position. Errorbars indicate 95\% Credible Intervals. }
    \label{fig:replies_gender_int}
\end{figure}

\textbf{Post Replies} Controlling for all other variables, we find women are predicted to have an 821\% increase in the rate of receiving replies to their posts compared to men (RR= 9.21, 95\% CrI[8.00, 10.28]) (see Figure \ref{fig:replies_gender_main}). For every 1 Standard Deviation (SD) increase in the average posts per week made by a politician there was a predicted decrease of 47\% in the rate of a post receiving replies (RR= 0.53, 95\% CrI[0.50, 0.57]). For every 1 SD point increase in the emotion score of a post, there was a predicted increase of 58\% in the rate of a post receiving replies (RR=1.58 , 95\% CrI[1.30, 1.95]). Political posts also predicted an increase of 88\% in the rate of a post receiving replies, compared to non-political posts (RR= 1.88 , 95\% CrI[1.68, 2.10]). Compared to normal speech posts, posts which contained hate speech had a 54\% increase in the rate of receiving replies (RR= 1.54, 95\% CrI[1.26, 1.88]), but we do not find fear speech predicted the rate of a post receiving replies (RR= 1.04, 95\% CrI[0.86, 1.27]). We observe that for men, those in the Cabinet (RR= 1.58 , 95\% CrI[1.15, 2.23]), the Governors (RR= 1.90 , 95\% CrI[1.26, 2.95]), the Chief Deputy of Staff (RR= 3.49, 95\% CrI[2.14, 5.87]), the head of the Federal Bureau of Investigation (RR= 6.11, 95\% CrI[5.16, 7.24]), the Vice President (RR= 6.82 , 95\% CrI[4.26, 11.59]), and the President (RR= 146.94, 95\% CrI[122.73, 174.16 ]), all had higher rates of receiving replies to their posts, compared to those in the House of Representatives. The greatest difference was observed for President Trump, who had an increase of 145.94\% in the rate of his posts receiving replies compared to those in the House of Representatives. In contrast, for the men in the Senate there is predicted a decrease of 55\% in the rate of a post receiving replies (RR= 0.45, 95\% CrI[0.35, 0.58]). For women in the Cabinet, we find that there is a 19\% smaller increase in  the rate at which their posts receive replies (RR=0.51 , 95\% CrI[0.35, 0.73]) compared to the men in the Cabinet. A similar pattern is observed for women in the Senate, there is a 97\% smaller increase in the rate of their posts receiving replies,  (RR= 0.05, 95\% CrI [0.04, 0.08]), compared to the men in the Senate. The interaction is also visualised in Figure \ref{fig:replies_gender_int}. Political positions without observations for both genders were excluded because model-based predictions for these cells rely entirely on extrapolation and are not informative for interpreting the interaction.


\begin{figure}[htbp] 
    \centering
    {\raggedright
    \hspace*{-0.6cm}%
    \includegraphics[width=0.40\textwidth]{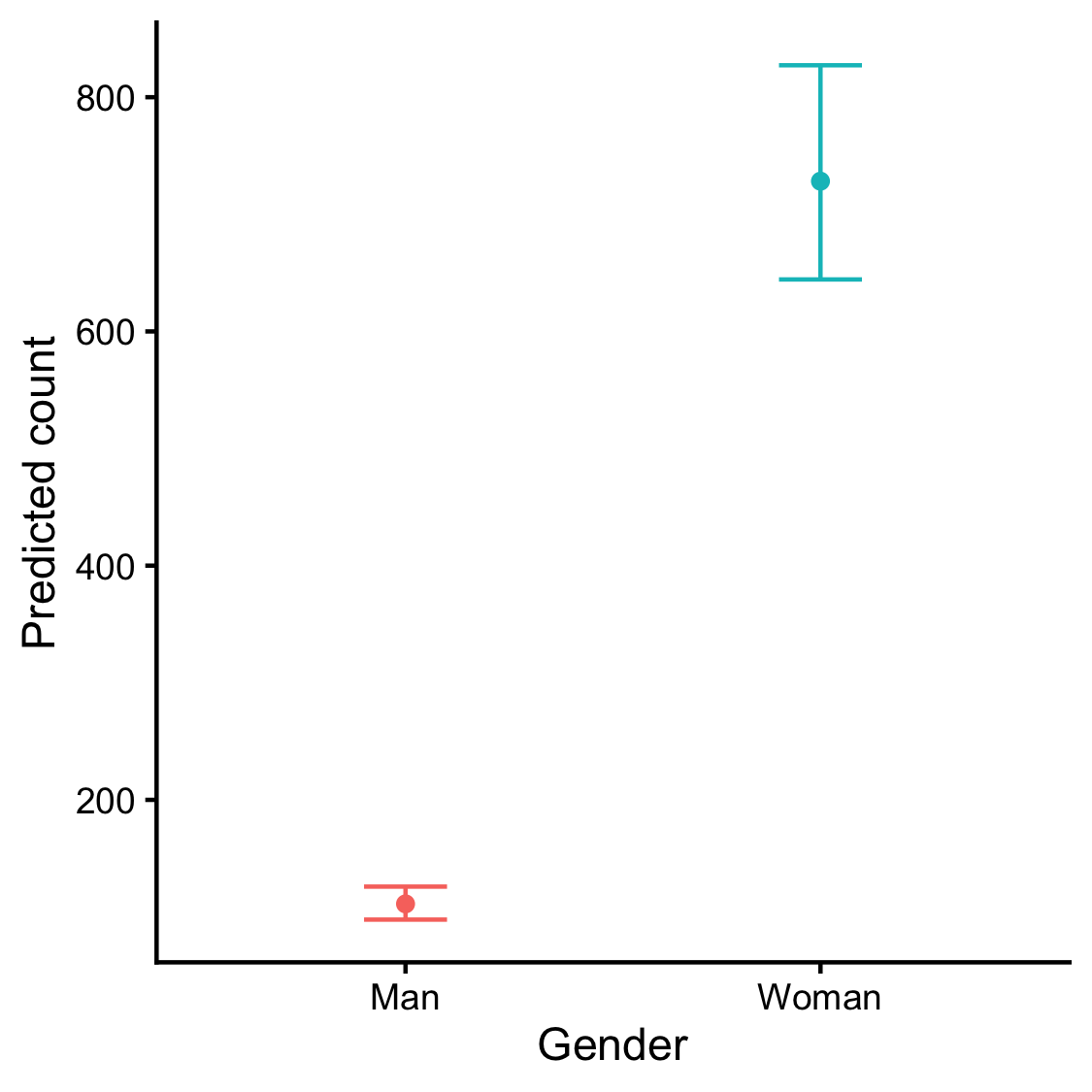}
    \par}
    \caption{Estimated counts of upvotes received by men and by women to their posts when controlling for covariates. Errorbars indicate 95\% Credible Intervals.}
    \label{fig:upvotes_gender_main}
\end{figure}

\begin{figure}[htbp] 
    \centering
    {\raggedright
    \hspace*{-0.5cm}%
    \includegraphics[width=0.5\textwidth]{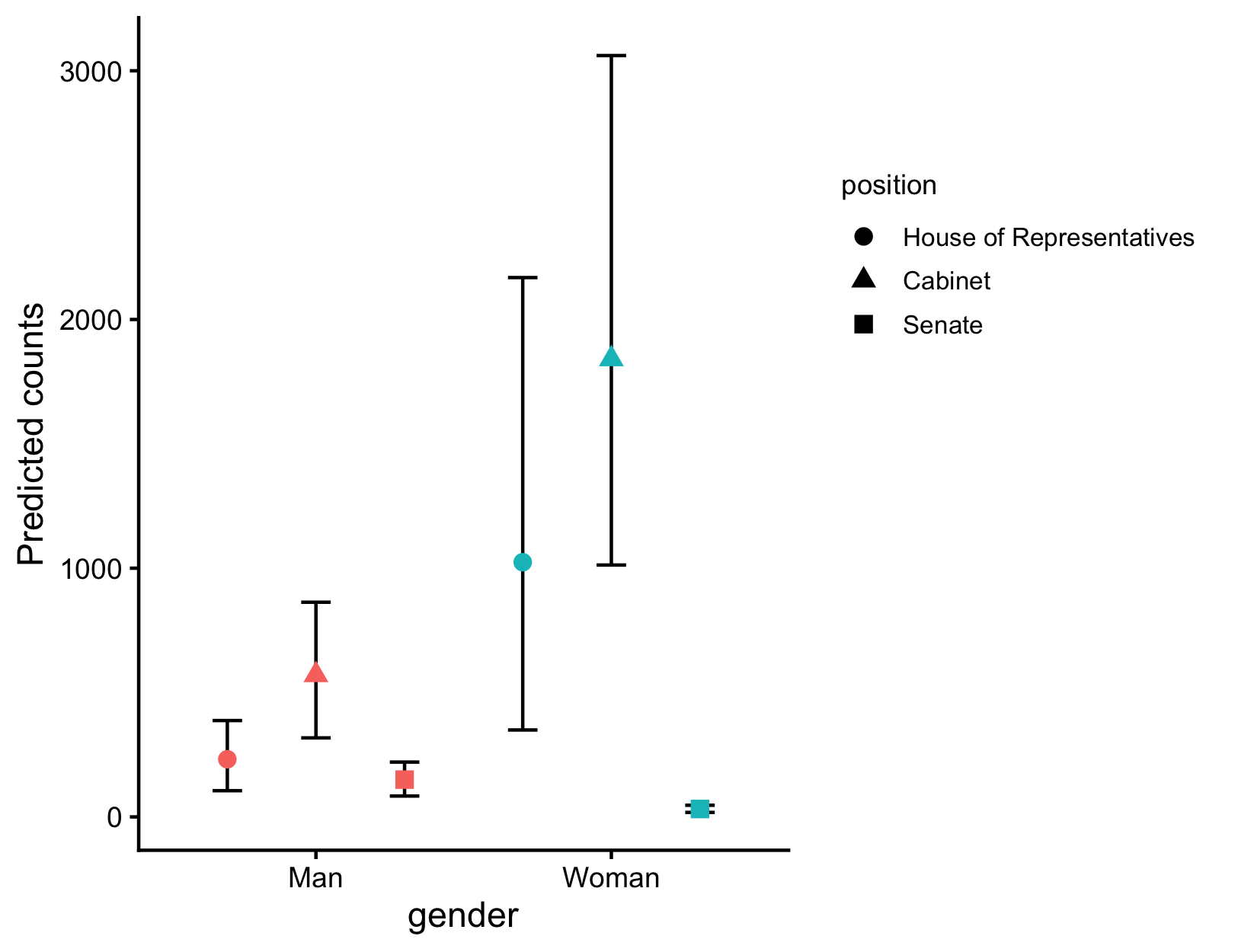}
    \par}
    \caption{Interaction of gender and political position for upvote counts. Estimates are predicted counts of replies received by posts of political elites depending on their gender and political position. Errorbars indicate 95\% Credible Intervals. }
    \label{fig:upvotes_gender_int}
\end{figure}

\textbf{Post Upvotes} Controlling for all other variables, women have a 555\% increase in the rate of receiving upvotes to their posts compared to men (RR=6.55 , 95\% CrI[5.76, 7.46]) (see Figure \ref{fig:upvotes_gender_main}). For every 1 Standard Deviation (SD) point increase in the average posts per week there is a predicted 38\% decrease in the rate that a post will receive upvotes (RR= 0.62, 95\% CrI[0.58, 0.66]). For every 1SD point increase in emotion score of a post, there is a predicted 21\% decrease in the rate that this post will receive upvotes (RR= 0.79, 95\% CrI[0.64, 0.97]). However, it is worth mentioning the upper bound of this coefficient is very close to an Rates Ratio of 1, which suggest that this association could be very small, so this result should be interpreted with caution. Compared to non-political posts, post which are political have a 63\% increase in the rate of receiving upvotes (RR= 1.63, 95\% CrI[1.46, 1.84]). Compared to normal speech posts, posts which contain hate speech have an increase of 40\% in the rate of receiving upvotes (RR= 1.40, 95\% CrI [1.13, 1.73]). In contrast, fear speech was not predictive of the rate of a post receiving upvotes (RR= 0.98 , 95\% CrI[0.81, 1.19]). Compared to men in the House of Representatives, those in the Cabinet (RR= 1.73, 95\% CrI[1.26, 2.48]), the Chief Deputy of Staff (RR= 3.78 , 95\% CrI[2.27, 6.55]). and the Head of the Federal Bureau of Investigation (RR= 6.69, 95\% CrI[5.53, 8.00]) were found to have increased rates of receiving upvotes to their posts.  Compared to the men in the Senate, the women had a 98\% smaller increase in the rate of receiving upvotes to their posts (RR= 0.04, 95\% CrI[0.03, 0.06]). The interaction is visualised in Figure \ref{fig:upvotes_gender_int}. To aid interpretation of the interaction, we report estimates only for political positions that are represented in the data for both genders.

\section{Discussion}
Our analysis presents a nuanced set of findings. Although we do not find gender differences in the linguistic composition of posts or in the use of hate speech, marked disparities emerge in emotional expression, content framing, the deployment of fear speech, and audience engagement. Taken together, these patterns suggest that content posted from the accounts of women political elites reflects communicative practices shaped by gendered expectations that remain present within this alt-tech environment.

We find that women political elites have lower odds of expressing anger than their male counterparts, but markedly higher odds of expressing joy. This pattern aligns with previous findings showing that Republican women politicians on Twitter tend to employ more joyful and fewer angry appeals, consistent with gendered expectations that women should minimise overt displays of aggression \cite{Russell03072023}. Our results therefore indicate that such expectations extend into Truth Social, despite its distinct ideological profile.

Women political elites are also significantly less likely to post political content than men, even when we control for posting frequency. This mirrors broader patterns of political engagement on social media, where men typically produce more overtly political material \cite{gender_normal_people_politics}. It also resonates with longstanding arguments concerning the Double Bind, in which women in public life must balance authority with approachability, yet face criticism if they appear either overly assertive or insufficiently warm \cite{jamieson1995beyond}. Within this context, the lower prevalence of political content in posts associated with women political elites on Truth Social may reflect a strategic presentation style oriented towards navigating these intersecting pressures.

Topic analysis reveals minimal substantive gender differences in the issues discussed by men and women political elites. Consistent with prior research showing minimal gender divergence in policy agendas and topics discussed by politicians on social media \cite{Russell02092021, Oden04052025}, both groups engage primarily with mainstream conservative themes, and we find no emphasis on issues traditionally associated with women. A similar pattern emerges in non-political content. Contrary to scholarship suggesting that women politicians often personalise their communication \cite{McGregor01022017, gender_diff_yt}, such motifs are largely absent from the topics identified here.

The absence of gender-coded issue emphasis across both political and non-political content is itself notable. One interpretation is that the Double Bind discourages overt engagement with traditionally feminine themes, a pattern consistent with previous findings that Republican women are less likely than Republican men to reference family life on social media\cite{family_difference_rep}. Another, not mutually exclusive, explanation is that the platform exerts strong pressure towards alignment with its dominant discursive agenda. Truth Social is characterised by high levels of ideological uniformity \cite{Gerard_Botzer_Weninger_2023}, consistent with research documenting increasing homogenisation within the Republican Party \cite{Thomsen2015RepublicanWomen, wineinger2022gendering}. Women political elites on the platform may therefore engage in this consolidated communicative space rather than differentiating themselves through gender-specific issues.

Our analysis also shows that posts associated with women political elites have substantially higher odds of containing fear speech, rhetoric that constructs an out-group as a threat to the in-group’s safety or values \cite{gerard2025bridgingnarrativedividecrossplatform}, compared to posts associated with men. Distinct from hate speech, which is explicitly derogatory, fear speech functions primarily through threat amplification. The elevated use of fear-oriented rhetoric among women political elites may reflect an alternative communicative pathway for participating in the platform’s alarmist discourse. Prior work shows that women politicians sometimes draw upon maternal or caregiving themes in their public communication \cite{McGregor01022017}. Within this context, fear speech may enable women to express vigilance towards perceived threats while avoiding the social penalties that often accompany more direct, confrontational or overtly hostile rhetoric from women. The topics associated with this rhetoric, such as references to fentanyl rather than ``illegal aliens" when discussing the border, suggest that fear speech may provide a gender-acceptable means of engaging in a politicized threat narrative.

Women political elites also receive substantially higher levels of audience engagement than their men counterparts, with greater rates of attracting both replies and upvotes. Importantly, this engagement advantage cannot be explained by rhetorical features associated with threat or emotion. Fear speech, despite being more prevalent in posts associated with women, does not predict increased engagement. Similarly, although anger is widely associated with increased engagement online \cite{anger_in_posts, biljaruzelska2022feelings}, women exhibit lower odds of using anger yet still receive markedly more replies and upvotes. These findings suggest that linguistic or affective features alone do not account for women’s engagement advantage.

Instead, the pattern is consistent with prior research showing that women politicians often attract higher user engagement across social media platforms \cite{Yarchi03072018, mcgregor2016talking}. One explanation advanced in this work is that women who achieve prominence in male-dominated political spaces often do so because they are required to demonstrate particularly high levels of competence, skill, or communicative effectiveness \cite{Yarchi03072018}. This dynamic may be especially pronounced on Truth Social, a platform whose user base is strongly associated with support for hegemonic masculinity and hostile sexism. Women who establish a meaningful presence within such an environment must negotiate both the general gendered constraints of political life \cite{Lawless_2009} and the more acute prejudices characteristic of this ideological community. Those who succeed under these conditions may therefore represent an especially select group, and their heightened engagement levels may reflect the cumulative filtering effects that shape which women appear, and are able to remain, on the platform.

\section{Limitations}
Several limitations should be acknowledged in this work. First, the time frame of our data collection may exclude significant events relevant to Donald Trump and Truth Social that occurred subsequently, such as debates over the release of the Epstein files. However, given our focus on broader communicative patterns rather than event-specific responses, and considering the size and temporal span of our dataset, the absence of particular events is unlikely to affect our central findings.

The topic modelling also inherently requires heuristic interpretation. The themes we present are our own analytical judgment and should be understood as interpretive rather than definitive. To facilitate transparency and enable future validation, detailed information about the topics has been included in Section B of the Appendix in the Supplementary Material. In addition, the high number of topics generated meant we did not have sufficient observations for each topic and therefore we could not consider the possible topic random effects on replies and upvotes count. To mitigate this, we controlled for content variance using the broader, robustly classified binary of political versus non-political content, which captures the primary axis of variance in audience engagement.

Additionally, due to political realities reflected in our sample (over 90\% are white political elites, see Section A in the Appendix in the Supplementary Material), we are unable to address questions about the interactive effects of gender and race on political communication patterns.

Lastly, as with any work which analyses naturalistic, observational data our statistical analyses cannot and should not be taken to show causal relationships between the predictors we found were credibly associated.

\section{Ethics}
This study was approved by the University of Edinburgh School of Informatics Ethics Committee [RT \#9516]. All analyses are observational, and any data shared as part of this work consist solely of information publicly available online from high-profile public figures.

\begin{acks}
This work was supported in part by the UKRI Centre for Doctoral Training in Designing Responsible Natural Language Processing, funded by the UKRI (Grant EP/Y030656/1),  the UKRI Centre for Doctoral Training in Natural Language Processing, funded by the UKRI (Grant EP/S022481/1), and the University of Edinburgh, School of Informatics.
\end{acks}

\section{Conclusion}
This study examined whether gendered patterns of elite political communication observed on mainstream social media persist within an alt-tech environment. Despite Truth Social’s ideological homogeneity and its association with hyper-masculine political norms, we find that many gendered dynamics closely mirror those documented on mainstream platforms, including differences in emotions expressed and audience engagement by gender. This is notable given that the platform’s political alignment might be expected to amplify such differences.

At the same time, some new distinctions are evident. Although men and women political elites engage with largely similar conservative themes, they differ in how these issues are framed and in their use of fear-oriented rhetoric. Future work should examine whether similar gendered patterns in fear-based appeals appear on mainstream platforms, helping to establish whether this difference is specific to Truth Social or reflective of broader gendered communication dynamics.

Overall, this work demonstrates that alt-tech platforms do not fundamentally reshape gendered political discourse. Instead, they largely reproduce dynamics observed on mainstream social media, with potentially selective adaptations shaped by platform-specific norms and audiences.

\bibliographystyle{ACM-Reference-Format}
\bibliography{references}

\appendix
\section{Appendix}
For the appendices, please refer to the Supplementary Materials available at the following link. This document provides detailed information on the training procedures, prompts employed, definitions, and the full results of the Bayesian analyses.

Supplementary Materials: \url{https://osf.io/wdxgy/overview?view_only=be843b57a7b649689d2e7b205bccbbd4}

\clearpage
\end{document}